\input harvmac
%\draftmode 
\noblackbox 
\input epsf

\newcount\figno
\figno=0
\def\fig#1#2#3{
\par\begingroup\parindent=0pt\leftskip=1cm\rightskip=1cm\parindent=0pt
\baselineskip=11pt \global\advance\figno by 1 \midinsert \epsfxsize=#3
\centerline{\epsfbox{#2}} \vskip 12pt {\bf 
Fig.\ \the\figno: } #1\par
\endinsert\endgroup\par
}
\def\figlabel#1{\xdef#1{\the\figno}}
\def\encadremath#1{\vbox{\hrule\hbox{\vrule\kern8pt\vbox{\kern8pt
\hbox{$\displaystyle #1$}\kern8pt} \kern8pt\vrule}\hrule}}

%%% special math symbols

\def\eqn#1#2{\xdef #1{(\secsym\the\meqno)}\writedef{#1\leftbracket#1}%
\global\advance\meqno by1$$#2\eqno#1\eqlabeL#1$$}

 %\SenMG  
\lref\SenMG{ A.~Sen, ``Non-BPS states and branes in string theory,'' 
arXiv:hep-th/9904207.  
%%CITATION = HEP-TH 9904207;%%
}

%\SenNU
\lref\SenNU{ A.~Sen, ``Rolling tachyon,'' JHEP {\bf 0204}, 048 (2002)
[arXiv:hep-th/0203211]. 
%%CITATION = HEP-TH 0203211;%%
}

%\SenIN
\lref\SenIN{ A.~Sen, ``Tachyon matter,'' JHEP {\bf 0207}, 065 (2002)
[arXiv:hep-th/0203265]. 
%%CITATION = HEP-TH 0203265;%%
}

\lref\GutperleAI{ M.~Gutperle and A.~Strominger, ``Spacelike branes,''
JHEP {\bf 0204}, 018 (2002) 
[arXiv:hep-th/0202210].
%%CITATION = HEP-TH 0202210;%%
}

%\SenAN
\lref\SenAN{ A.~Sen, ``Field theory of tachyon matter,'' Mod.\ Phys.\
Lett.\ A {\bf 17}, 1797 (2002) 
[arXiv:hep-th/0204143].
%%CITATION = HEP-TH 0204143;%%
}

%\SenQA
\lref\SenQA{ A.~Sen, ``Time and tachyon,'' arXiv:hep-th/0209122.
%%CITATION = HEP-TH 0209122;%%
}

%\StromingerPC
\lref\StromingerPC{ A.~Strominger, ``Open string creation by
S-branes,'' arXiv:hep-th/0209090. 
%%CITATION = HEP-TH 0209090;%%
}

%\GutperleXF
\lref\GutperleXF{ M.~Gutperle and A.~Strominger, ``Timelike boundary
Liouville theory,'' Phys.\ Rev.\ D {\bf 67}, 
126002 (2003) [arXiv:hep-th/0301038].
%%CITATION = HEP-TH 0301038;%%
}

%\MaloneyCK
\lref\MaloneyCK{ A.~Maloney, A.~Strominger and X.~Yin, ``S-brane
thermodynamics,'' arXiv:hep-th/0302146. 
%%CITATION = HEP-TH 0302146;%%
}

\lref\KarczmarekXM{ J.~L.~Karczmarek, H.~Liu, J.~Maldacena and
A.~Strominger, ``UV finite brane decay,'' 
arXiv:hep-th/0306132. 
%%CITATION = HEP-TH 0306132;%%
}

\lref\LarsenWC{ F.~Larsen, A.~Naqvi and S.~Terashima, ``Rolling
tachyons and decaying branes,'' JHEP {\bf 0302}, 
039 (2003) [arXiv:hep-th/0212248].
%%CITATION = HEP-TH 0212248;%%
}

%\ConstableRC
\lref\ConstableRC{ N.~R.~Constable and F.~Larsen, ``The rolling
tachyon as a matrix model,'' JHEP {\bf 0306}, 017 
(2003) [arXiv:hep-th/0305177].
%%CITATION = HEP-TH 0305177;%%
}

%\ChenFP
\lref\ChenFP{ B.~Chen, M.~Li and F.~L.~Lin, ``Gravitational radiation
of rolling tachyon,'' JHEP {\bf 0211}, 050 
(2002) [arXiv:hep-th/0209222].
%%CITATION = HEP-TH 0209222;%%
}

%\OkudaYD
\lref\OkudaYD{ T.~Okuda and S.~Sugimoto, ``Coupling of rolling tachyon
to closed strings,'' Nucl.\ Phys.\ B {\bf 
647}, 101 (2002) [arXiv:hep-th/0208196].
%%CITATION = HEP-TH 0208196;%%
}

%\LambertZR
\lref\LambertZR{ N.~Lambert, H.~Liu and J.~Maldacena, ``Closed strings
from decaying D-branes,'' 
arXiv:hep-th/0303139.
%%CITATION = HEP-TH 0303139;%%
}

%\GaiottoRM
\lref\GaiottoRM{ D.~Gaiotto, N.~Itzhaki and L.~Rastelli, ``Closed
strings as imaginary D-branes,'' 
arXiv:hep-th/0304192.
%%CITATION = HEP-TH 0304192;%%
}

%\MukhopadhyayEN
\lref\MukhopadhyayEN{ P.~Mukhopadhyay and A.~Sen, ``Decay of unstable
D-branes with electric field,'' JHEP {\bf 
0211}, 047 (2002) [arXiv:hep-th/0208142].
%%CITATION = HEP-TH 0208142;%%
}

%\ReyXS
\lref\ReyXS{ S.~J.~Rey and S.~Sugimoto, ``Rolling tachyon with
electric and magnetic fields: T-duality approach,'' 
Phys.\ Rev.\ D {\bf 67}, 086008 (2003) [arXiv:hep-th/0301049].
%%CITATION = HEP-TH 0301049;%%
}

%\MoellerVX
\lref\MoellerVX{ N.~Moeller and B.~Zwiebach, ``Dynamics with
infinitely many time derivatives and rolling 
tachyons,'' JHEP {\bf 0210}, 034 (2002) [arXiv:hep-th/0207107].
%%CITATION = HEP-TH 0207107;%%
}

%\SugimotoFP
\lref\SugimotoFP{ S.~Sugimoto and S.~Terashima, ``Tachyon matter in
boundary string field theory,'' JHEP {\bf 
0207}, 025 (2002) [arXiv:hep-th/0205085].
%%CITATION = HEP-TH 0205085;%%
}

\lref\Maldacena{J. Maldacena, I. Klebanov and N. Seiberg, Talks at
Strings 2003} 

%\KlusonTE
\lref\KlusonTE{ J.~Kluson, ``Time dependent solution in open bosonic string field theory,'' arXiv:hep-th/0208028.
%%CITATION = HEP-TH 0208028;%%
}

%\KlusonRD
\lref\KlusonRD{ J.~Kluson, ``Particle production on half S-brane,''
arXiv:hep-th/0306002. 
%%CITATION = HEP-TH 0306002;%%
}

%\IshidaCJ
\lref\IshidaCJ{ A.~Ishida and S.~Uehara, ``Rolling down to D-brane and
tachyon matter,'' JHEP {\bf 0302}, 050 
(2003) [arXiv:hep-th/0301179].
%%CITATION = HEP-TH 0301179;%%
}

%\SchomerusVV
\lref\SchomerusVV{ V.~Schomerus, ``Rolling tachyons from Liouville
theory,'' arXiv:hep-th/0306026. 
%%CITATION = HEP-TH 0306026;%%
}

%\MoellerGG
\lref\MoellerGG{ N.~Moeller and M.~Schnabl, ``Tachyon condensation in
open-closed p-adic string theory,'' 
arXiv:hep-th/0304213.
%%CITATION = HEP-TH 0304213;%%
}

%\FujitaEX
\lref\FujitaEX{ M.~Fujita and H.~Hata, ``Time dependent solution in
cubic string field theory,'' JHEP {\bf 0305}, 
043 (2003) [arXiv:hep-th/0304163].
%%CITATION = HEP-TH 0304163;%%
}

%\KutasovER
\lref\KutasovER{ D.~Kutasov and V.~Niarchos, ``Tachyon effective
actions in open string theory,'' 
arXiv:hep-th/0304045.
%%CITATION = HEP-TH 0304045;%%
}

%\StromingerFN
\lref\StromingerFN{ A.~Strominger and T.~Takayanagi, ``Correlators in
timelike bulk Liouville theory,'' 
arXiv:hep-th/0303221.
%%CITATION = HEP-TH 0303221;%%
}

%\ChenYQ
\lref\ChenYQ{ C.~M.~Chen, D.~V.~Gal'tsov and M.~Gutperle, ``S-brane
solutions in supergravity theories,'' Phys.\ 
Rev.\ D {\bf 66}, 024043 (2002) [arXiv:hep-th/0204071].
%%CITATION = HEP-TH 0204071;%%
}

%\KruczenskiAP
\lref\KruczenskiAP{ M.~Kruczenski, R.~C.~Myers and A.~W.~Peet,
``Supergravity S-branes,'' JHEP {\bf 0205}, 039 
(2002) [arXiv:hep-th/0204144].
%%CITATION = HEP-TH 0204144;%%
}

%\BuchelTJ
\lref\BuchelTJ{ A.~Buchel, P.~Langfelder and J.~Walcher, ``Does the
tachyon matter?,'' Annals Phys.\  {\bf 302}, 
78 (2002) [arXiv:hep-th/0207235].
%%CITATION = HEP-TH 0207235;%%
}

\lref\GibbonsMD{ G.~W.~Gibbons, ``Cosmological evolution of the
rolling tachyon,'' Phys.\ Lett.\ B {\bf 537}, 1 
(2002) [arXiv:hep-th/0204008].
%%CITATION = HEP-TH 0204008;%%
}

%\AlexandrovNN
\lref\AlexandrovNN{
S.~Y.~Alexandrov, V.~A.~Kazakov and D.~Kutasov,
``Non-perturbative effects in matrix models and D-branes,''
arXiv:hep-th/0306177.
%%CITATION = HEP-TH 0306177;%%
}

%\FrolovRR
\lref\FrolovRR{ A.~V.~Frolov, L.~Kofman and A.~A.~Starobinsky,
``Prospects and problems of tachyon matter 
cosmology,'' Phys.\ Lett.\ B {\bf 545}, 8 (2002) [arXiv:hep-th/0204187].
%%CITATION = HEP-TH 0204187;%%
}

%\KlebanovQA
\lref\KlebanovQA{ I.~R.~Klebanov, ``String theory in two-dimensions,''
arXiv:hep-th/9108019. 
%%CITATION = HEP-TH 9108019;%%
}

%\GinspargIS
\lref\GinspargIS{ P.~Ginsparg and G.~W.~Moore, ``Lectures On 2-D
Gravity And 2-D String Theory,'' 
arXiv:hep-th/9304011.
%%CITATION = HEP-TH 9304011;%%
}

%\PolchinskiMB
\lref\PolchinskiMB{ J.~Polchinski, ``What is string theory?,''
arXiv:hep-th/9411028. 
%%CITATION = HEP-TH 9411028;%%
}

%\McGreevyKB
\lref\McGreevyKB{ J.~McGreevy and H.~Verlinde, ``Strings from
tachyons: The c = 1 matrix reloated,'' 
arXiv:hep-th/0304224.
%%CITATION = HEP-TH 0304224;%%
}

%\McGreevyEP
\lref\McGreevyEP{ J.~McGreevy, J.~Teschner and H.~Verlinde,
``Classical and quantum D-branes in 2D string 
theory,'' arXiv:hep-th/0305194.
%%CITATION = HEP-TH 0305194;%%
}

%\DouglasUP
\lref\DouglasUP{ M.~R.~Douglas, I.~R.~Klebanov, D.~Kutasov,
J.~Maldacena, E.~Martinec and N.~Seiberg, ``A new hat 
for the c = 1 matrix model,'' arXiv:hep-th/0307195.
%%CITATION = HEP-TH 0307195;%%
}

%\KlebanovKM
\lref\KlebanovKM{ I.~R.~Klebanov, J.~Maldacena and N.~Seiberg,
``D-brane decay in two-dimensional string theory,'' 
JHEP {\bf 0307}, 045 (2003) [arXiv:hep-th/0305159].
%%CITATION = HEP-TH 0305159;%%
}

%\TakayanagiSM
\lref\TakayanagiSM{ T.~Takayanagi and N.~Toumbas, ``A matrix model
dual of type 0B string theory in two 
dimensions,'' arXiv:hep-th/0307083.
%%CITATION = HEP-TH 0307083;%%
}

%\ZamolodchikovAH
\lref\ZamolodchikovAH{ A.~B.~Zamolodchikov and A.~B.~Zamolodchikov,
``Liouville field theory on a pseudosphere,'' 
arXiv:hep-th/0101152.
%%CITATION = HEP-TH 0101152;%%
}

%\FateevIK
\lref\FateevIK{ V.~Fateev, A.~B.~Zamolodchikov and
A.~B.~Zamolodchikov, ``Boundary Liouville field theory. I: 
Boundary state and boundary two-point function,'' arXiv:hep-th/0001012.
%%CITATION = HEP-TH 0001012;%%
}

%\MartinecKA
\lref\MartinecKA{ E.~J.~Martinec, ``The annular report on non-critical
string theory,'' arXiv:hep-th/0305148. 
%%CITATION = HEP-TH 0305148;%%
}

%\DiFrancescoUD
\lref\DiFrancescoUD{ P.~Di Francesco and D.~Kutasov, ``World sheet and
space-time physics in two-dimensional 
(Super)string theory,'' Nucl.\ Phys.\ B {\bf 375}, 119 (1992) [arXiv:hep-th/9109005].
%%CITATION = HEP-TH 9109005;%%
}

%\ShenkerUF
\lref\ShenkerUF{ S.~H.~Shenker, ``The Strength Of Nonperturbative
Effects In String Theory,'' RU-90-47 
%\href{http://www.slac.stanford.edu/spires/find/hep/www?r=ru-90-47}{SPIRES
%entry}
{\it Presented at the Cargese Workshop on Random Surfaces, Quantum
Gravity and Strings, Cargese, France, May 28 - 
Jun 1, 1990} }

%\PolchinskiMT
\lref\PolchinskiMT{ J.~Polchinski,
%``Dirichlet-Branes and Ramond-Ramond Charges,''
Phys.\ Rev.\ Lett.\  {\bf 75}, 4724 (1995) [arXiv:hep-th/9510017].
%%CITATION = HEP-TH 9510017;%%
}

%\ConstableRC
\lref\ConstableRC{
N.~R.~Constable and F.~Larsen,
``The rolling tachyon as a matrix model,''
JHEP {\bf 0306}, 017 (2003)
[arXiv:hep-th/0305177].
%%CITATION = HEP-TH 0305177;%%
}

%\LarsenWC
\lref\LarsenWC{
F.~Larsen, A.~Naqvi and S.~Terashima,
`Rolling tachyons and decaying branes,''
JHEP {\bf 0302}, 039 (2003)
[arXiv:hep-th/0212248].
%%CITATION = HEP-TH 0212248;%%
}

%\PolchinskiUQ
\lref\PolchinskiUQ{ J.~Polchinski, ``Classical Limit Of
(1+1)-Dimensional String Theory,'' Nucl.\ Phys.\ B {\bf 
362}, 125 (1991).
%%CITATION = NUPHA,B362,125;%%
}

%\MooreGB
\lref\MooreGB{ G.~W.~Moore and R.~Plesser, ``Classical scattering in
(1+1)-dimensional string theory,'' Phys.\ 
Rev.\ D {\bf 46}, 1730 (1992) [arXiv:hep-th/9203060].
%%CITATION = HEP-TH 9203060;%%
}

%\DasKA
\lref\DasKA{ S.~R.~Das and A.~Jevicki, ``String Field Theory And
Physical Interpretation Of D = 1 Strings,'' Mod.\ 
Phys.\ Lett.\ A {\bf 5}, 1639 (1990).
%%CITATION = MPLAE,A5,1639;%%
}

%\BrusteinBI
\lref\BrusteinBI{ R.~Brustein and B.~A.~Ovrut, ``Nonperturbative
Effects In 2-D String Theory Or Beyond The 
Liouville Wall,'' Int.\ J.\ Mod.\ Phys.\ A {\bf 12}, 3477 (1997)
[arXiv:hep-th/9209081]. 
%%CITATION = HEP-TH 9209081;%%
}

%\BrusteinMY
\lref\BrusteinMY{ R.~Brustein, M.~Faux and B.~A.~Ovrut, ``Instanton
effects in matrix models and string effective 
Lagrangians,'' Nucl.\ Phys.\ B {\bf 433}, 67 (1995) [arXiv:hep-th/9406179].
%%CITATION = HEP-TH 9406179;%%
}

%\JevickiQN
\lref\JevickiQN{
A.~Jevicki,
``Development in 2-d string theory,''
arXiv:hep-th/9309115.
%%CITATION = HEP-TH 9309115;%%
}

%\PolyakovQX
\lref\PolyakovQX{
A.~M.~Polyakov,
``Selftuning Fields And Resonant Correlations In 2-D Gravity,''
Mod.\ Phys.\ Lett.\ A {\bf 6}, 635 (1991).
%%CITATION = MPLAE,A6,635;%%
}

%\DiFrancescoSS
\lref\DiFrancescoSS{
P.~Di Francesco and D.~Kutasov,
``Correlation functions in 2-D string theory,''
Phys.\ Lett.\ B {\bf 261}, 385 (1991);
%%CITATION = PHLTA,B261,385;%%
%}
%\DiFrancescoUD
%\lref\DiFrancescoUD{
%P.~Di Francesco and D.~Kutasov,
``World sheet and space-time physics in two-dimensional (Super)string theory,''
Nucl.\ Phys.\ B {\bf 375}, 119 (1992)
[arXiv:hep-th/9109005].
%%CITATION = HEP-TH 9109005;%%
}

%\MarcusVS
\lref\MarcusVS{
N.~Marcus,
``Unitarity And Regularized Divergences In String Amplitudes,''
Phys.\ Lett.\ B {\bf 219}, 265 (1989).
%%CITATION = PHLTA,B219,265;%%
}

%\CrapsJP
\lref\CrapsJP{
B.~Craps, P.~Kraus and F.~Larsen,
``Loop corrected tachyon condensation,''
JHEP {\bf 0106}, 062 (2001)
[arXiv:hep-th/0105227].
%%CITATION = HEP-TH 0105227;%%
}

%\WittenZW
\lref\WittenZW{
E.~Witten,
``Anti-de Sitter space, thermal phase transition, and confinement in  gauge theories,''
Adv.\ Theor.\ Math.\ Phys.\  {\bf 2}, 505 (1998)
[arXiv:hep-th/9803131].
%%CITATION = HEP-TH 9803131;%%
}

%\GaiottoYF
\lref\GaiottoYF{
D.~Gaiotto, N.~Itzhaki and L.~Rastelli,
``On the BCFT description of holes in the c = 1 matrix model,''
arXiv:hep-th/0307221.
%%CITATION = HEP-TH 0307221;%%
}

%\KazakovPM
\lref\KazakovPM{
V.~Kazakov, I.~K.~Kostov and D.~Kutasov,
``A matrix model for the two-dimensional black hole,''
Nucl.\ Phys.\ B {\bf 622}, 141 (2002)
[arXiv:hep-th/0101011].
%%CITATION = HEP-TH 0101011;%%
}

\lref\SenguptaBT{
A.~M.~Sengupta and S.~R.~Wadia,
``Excitations And Interactions In D = 1 String Theory,''
Int.\ J.\ Mod.\ Phys.\ A {\bf 6}, 1961 (1991).
%%CITATION = IMPAE,A6,1961;%%
}

\lref\GrossST{
D.~J.~Gross and I.~R.~Klebanov,
``Fermionic String Field Theory Of C = 1 Two-Dimensional Quantum Gravity,''
Nucl.\ Phys.\ B {\bf 352}, 671 (1991).
%%CITATION = NUPHA,B352,671;%%
}

\lref\MandalUA{
G.~Mandal, A.~M.~Sengupta and S.~R.~Wadia,
``Interactions and scattering in d = 1 string theory,''
Mod.\ Phys.\ Lett.\ A {\bf 6}, 1465 (1991).
%%CITATION = MPLAE,A6,1465;%%
}

\lref\DharCS{
A.~Dhar, G.~Mandal and S.~R.~Wadia,
``A Time dependent classical solution of c = 1 string field theory and
nonperturbative effects,'' 
Int.\ J.\ Mod.\ Phys.\ A {\bf 8}, 3811 (1993)
[arXiv:hep-th/9212027].
%%CITATION = HEP-TH 9212027;%%
}

\lref\DharHR{
A.~Dhar, G.~Mandal and S.~R.~Wadia,
``Nonrelativistic fermions, coadjoint orbits of W(infinity) and string field theory at c = 1,''
Mod.\ Phys.\ Lett.\ A {\bf 7}, 3129 (1992)
[arXiv:hep-th/9207011].
%%CITATION = HEP-TH 9207011;%%
}

\def\lambdah{\hat{\lambda}}

\def\p{\partial}

\def\mb{{\mu}}

\Title {\vbox{ \baselineskip12pt
\hbox{hep-th/0308047}\hbox{UCLA-03-TEP-22}}} {\vbox{
\centerline{D-brane Dynamics in the $c=1$  Matrix Model     }  }} 

\centerline{ Michael Gutperle\foot{gutperle@physics.ucla.edu} and Per
Kraus\foot{pkraus@physics.ucla.edu}}

\bigskip
\centerline{ Department of Physics and Astronomy, UCLA, Los Angeles,
CA 90095-1547, USA} 
\smallskip

\vskip .3in \centerline{\bf Abstract} 
\baselineskip15pt
\vskip.1cm Recent work has shown that unstable D-branes in
two dimensional string theory are represented by eigenvalues in a dual
matrix model.  We elaborate on this 
proposal by showing how to systematically include higher order effects
in string perturbation theory.   The full 
closed string state produced by a rolling open string tachyon
corresponds to a sum of string amplitudes with any 
number of boundaries and closed string vertex operators. These
contributions are easily extracted from the matrix 
model. As in the AdS/CFT correspondence, the sum of planar diagrams in
the open string theory is directly related to 
  the classical theory in the bulk, i.e. sphere diagrams.  We also
comment on the description of static 
  D-branes in the matrix model, in terms of a solution representing a
deformed Fermi sea. 
\smallskip
\Date{}

\baselineskip14pt

\newsec{Introduction}

The presence of a tachyon in the spectrum of a string theory signals
an instability of the string background. 
Recently, following the seminal work of Sen \SenMG, a great deal of
progress has been made in understanding the 
dynamics of open string tachyon condensation, which is associated with
the decay of unstable D-branes. In 
particular the rolling tachyon solution  \SenNU\SenIN\ provides  an
exact boundary CFT describing the decay of a 
bosonic D-brane. The rolling tachyon has been analyzed from various
points of view, among them are  open string 
particle production in the rolling tachyon solution \refs{
\StromingerPC,\GutperleXF,\KlusonRD}, supergravity 
description of S-branes
\refs{\GutperleAI,\ChenYQ,\KruczenskiAP,\BuchelTJ}, time dependent
solutions in string 
field theory
\refs{\MoellerVX,\KlusonTE,\SugimotoFP,\LarsenWC,
\MoellerGG,\FujitaEX,\KutasovER,\ConstableRC},
thermodynamics 
\refs{\MaloneyCK}, rolling tachyons in different backgrounds
\refs{\MukhopadhyayEN,\ReyXS,\KarczmarekXM}, 
radiation into closed strings \refs{\ChenFP,\OkudaYD, \LambertZR},
application to of rolling tachyons to cosmology 
\refs{\SenAN,\SenQA, \GibbonsMD,\FrolovRR} and generalizations to
closed string tachyon condensation 
\refs{\StromingerFN,\SchomerusVV}.

In \LambertZR\ it was found that to leading order in $g_s$ the energy
radiated into closed strings from the decay 
of unstable D0 branes diverges, suggesting that the unstable brane
decays completely into closed strings.  An interesting question is
how to reconcile this with a weak coupling open string analysis, which 
indicates a decay into a new form of ``tachyon matter'' \SenIN.  

Strings in two dimensions are interesting toy models to analyze
questions which are difficult to address in 
critical string theory, since one can take advantage of a reformulation
of the theory as a hermitian 
matrix model in the double scaling limit (See
\refs{\KlebanovQA,\GinspargIS, \PolchinskiMB,\JevickiQN} and references 
therein). In the singlet sector the degrees of freedom  reduce to the
matrix eigenvalues,  whose dynamics is in  turn
equivalent to a theory of
free fermions in a potential. In the double scaling limit, the 
eigenvalue distribution becomes continuous and 
is interpreted as a spatial dimension. 

The analysis of the decay of an unstable brane in two dimensional
string theory was initiated in \McGreevyKB, 
where it was proposed that in the free fermion formulation of the two
dimensional string an unstable brane 
corresponds to a free fermion which is moved from the Fermi sea to top
of the inverted harmonic oscillator  
potential. The subsequent decay is described by the rolling of the
fermion from the top to the spatial region occupied by the Fermi sea.

In \KlebanovKM\ (see also \McGreevyEP\MartinecKA) this matrix model
process was identified with a rolling tachyon 
boundary CFT which is constructed by tensoring Sen's rolling tachyon
in the time direction with a boundary state 
for the Liouville theory introduced by the Zamolodchikovs
\ZamolodchikovAH\FateevIK. This boundary state 
corresponds to a Dirichlet brane localized in the Liouville direction
and has a tachyon in the open string 
spectrum.

The nonrelativistic fermion that is rolling down the potential becomes
relativistic at late times. The closed  
string excitations can be identified with the boson which appears in
the bosonization of this fermion. This 
naturally leads to an identification of the rolling fermion at late
times with the coherent state produced by the 
closed string radiation at leading order in string perturbation
theory. An important check of this proposal 
\KlebanovKM\ is the agreement of the outgoing radiation derived from
the matrix model and the disk amplitudes in 
the two dimensional string theory.

The matrix model  is a very convenient and powerful method
for computing amplitudes in string perturbation theory.  We will put to use 
old results on the tree level tachyon S-matrix to extend the study of the
rolling tachyon to higher order in $g_s$.     The higher order contributions 
to the outgoing closed string state can then be identified as coming from 
worldsheets with various numbers of holes and vertex operators.   Using the
closed string tree level S-matrix, we  will sum up the contributions from 
planar diagrams with arbitrary numbers of holes, just as in
AdS/CFT. Indeed, it is 
now appreciated that the relation between
two dimensional string theory and the matrix model is perhaps 
the simplest realization of a
holographic duality.  Beyond the disk one-point function there are no
continuum calculations we can compare our results against.  However,
we can make 
some consistency checks, such as the fact that introducing $N$ unstable
D-branes correctly yields a factor of $N$ for each worldsheet boundary.

The plan for this note is as follows. Section two reviews  the
computation of the S-matrix for the scattering of closed string
tachyons in the matrix model, following Polchinski \PolchinskiUQ\PolchinskiMB.
In section three the fermionic description of the decay of the
unstable D-brane \McGreevyKB\KlebanovKM\ is reviewed.
 After bosonization, the relation between
in and out bosonic oscillators is used to derive the 
S-matrix for closed strings in the decaying brane background. 
In section four these matrix model results are used to predict the 
disk n-point function in the rolling tachyon background. 
In section five the results are compared to string perturbation theory
and it is found that worldsheets with multiple  boundaries 
contribute once  operator ordering is taken into account. 
In section six a modification of the matrix model is
discussed where classically 
the fermion sits at the top of the potential forever. It is
suggested that this state corresponds to the unstable D0 brane with
the tachyon set to zero.  We test this proposal by studying
 closed string scattering in
this background.   
We close this note with a discussion and speculation regarding our
results.

\newsec{Review of Scattering in the Matrix Model}

The $c=1$ matrix model has a spacetime interpretation consisting
of massless particles (``tachyons'') propagating in an inhomogeneous
$1+1$ dimensional spacetime.  The spacetime is effectively semi-infinite
due to the presence of an exponentially rising tachyon condensate (the
``Liouville wall''),  and one  computes an S-matrix describing the
 scattering  of tachyons.  While only
limited results are available in the usual worldsheet CFT approach 
\refs{\PolyakovQX,\DiFrancescoSS}
the matrix model formulation leads to explicit results to all orders in the
string coupling.   Tree level amplitudes are particularly simple to
extract, as we now review.  We mainly follow the discussion in
\PolchinskiMB.  Additional reviews include 
\refs{\KlebanovQA,\GinspargIS,\JevickiQN,\MandalUA}.

Starting from an action for $N\times N$ hermitian matrices,
\eqn\za{ S=\beta N \int\! dt \left\{\half \Tr (\dot{M})^2 - \Tr
V(M) \right\}}
a standard procedure (see \PolchinskiMB)
 leads to a second quantized Hamiltonian describing
nonrelativistic fermions moving in an inverted harmonic oscillator
potential,
\eqn\zb{H = \int \! dx \left\{
\half \p_x \psi^\dagger \p_x \psi - {x^2 \over 2}
\psi^\dagger \psi + \mu \psi^\dagger \psi \right\}.}
$\psi$ is a fermionic field obeying the anticommutation relations
\eqn\zc{\{\psi(x,t),\psi^\dagger(x',t)\}= \delta(x-x').}
We will be interested in the region $x \leq 0$; the role of the
second $x>0$ region is explained in \refs{\TakayanagiSM,\DouglasUP}.
 In the  ground state
of our system all energy levels up to $E=0$ are filled by fermions,
so the Fermi surface is a distance $\mu$ below the top of the
potential.  Classically, fermions at the Fermi surface reflect
off the potential at $x= -\sqrt{2\mu}$.   $\mu$ is related to the
string coupling:
\eqn\zd{ \mu \sim 1/g.}

One class of excited states consists of smooth fluctuations of the
Fermi surface.  Fluctuations propagating in from $x =
-\infty$ scatter off the potential and propagate back out to
 $x =  -\infty$.  This is the matrix model version of
the tachyon S-matrix.   The tree level S-matrix is obtained by
treating the fermions classically.  In the classical limit, each
fermion is described by a point in phase space, moving according to
its classical equations of motion,
and we think of the Fermi sea as the region of phase space occupied by
fermions.  Since the single particle Hamiltonian is
\eqn\ze{H= {p^2 -x^2 \over 2}+\mu,}
filling states up to $E=0$ means that in the ground state the Fermi
sea is the region bounded by $p_\pm(x)_{gnd}$, 
where
\eqn\zf{ p_\pm(x)_{gnd}= \pm \sqrt{x^2-2\mu}.}
Excited states are then described by other choices for $p_\pm(x)$, and
have a total energy (we drop the constant $N\mu$ term)
\eqn\zg{\eqalign{ H &= {
1 \over 2\pi} \int_{-\infty}^\infty \! dx \int_{p_-}^{p_+}
\! dp ~\half (p^2-x^2) \cr
 &={1 \over 2\pi} \int_{-\infty}^\infty \! dx \left\{
{1 \over 6}(p_+^3-p_-^3)-\half x^2(p_+-p_-)\right\}.}}
It is convenient to write
\eqn\zh{ p_\pm(x,t) = \mp x \pm {1 \over x} \epsilon_{\pm}(x,t)}
and to further relate $\epsilon_\pm$ to a massless scalar field $S$  via
\eqn\zha{\pi^{-1/2} \epsilon_\pm(q,t) = \pm \pi_S(q,t)-\p_q S(q,t)}
where
\eqn\zhb{q = -\ln(-x).}
The Hamiltonian \zg\ then becomes
\eqn\zhc{  H = \half \int_{-\infty}^\infty \! dq
\left\{ \pi_S^2 +(\p_q S)^2 + \pi^{1/2} e^{2q} \left[
 \p_q S \pi_S^2 + {1 \over 3} (\p_q S)^3\right] \right\}}
and
\eqn\zhd{[S(q,t),\pi_S(q',t)]=i\delta(q-q').}
For $q\rightarrow -\infty$, $S$ becomes a free massless scalar field,
admitting the mode expansion
\eqn\zhe{S(q,t)=\int_{-\infty}^\infty \!
{dk \over \sqrt{8\pi^2 k^2}} \left[a_k e^{-i|k|t+ikq} + a_k^\dagger
e^{i|k|t -ikq}\right]}
with
\eqn\zhf{[a_k,a^\dagger_{k'}]=2\pi |k| \delta(k-k').}
Due to  reflection off the wall, only half of the operators $a_k$
are independent.  Finding the relation between the rightmovers, $a_{k>0}$, and
the leftmovers, $a_{k<0}$,
 is equivalent to computing the S-matrix for scattering
from the wall.  This can be achieved as follows \PolchinskiMB\MooreGB.

Given the classical equations of motion
\eqn\zi{ \dot{x} = p, \quad \dot{p} = x, }
we have the conserved quantities
\eqn\zj{ v= (-x-p)e^{-t}, \quad w= (-x+p)e^t,}
as well as arbitrary powers of these. Scattering amplitudes follow
from equating the values of the conserved quantities at early and
late times.  In particular, we consider the conserved quantities
\eqn\zk{  v_{mn} = e^{(n-m)t}\int_{F-F_0} {dp \, dx \over 2\pi}
(-x-p)^m (-x+p)^n}
where we subtract off the static Fermi $F_0$ sea for finiteness. At
early times the rightmoving fluctuations are 
$\epsilon_+(q,t)=\epsilon_+(t-q)$, and at late times the leftmoving
fluctuations are 
$\epsilon_-(q,t)=\epsilon_-(t+q)$.   The conserved quantities \zk\
then become, after a short calculation 
\eqn\zl{\eqalign{ v_{mn} & =
   {2^n \over 2\pi(m+1)}\int_{-\infty}^\infty \! dt\,
e^{(n-m)(t-q)} \left[
(\epsilon_+(t-q))^{m+1} - \mu^{m+1}\right] \cr
 &= {2^m \over 2\pi(n+1)}\int_{-\infty}^\infty \! dt\,
e^{(n-m)(t+q)} \left[
(\epsilon_-(t+q))^{n+1} - \mu^{n+1}\right].
}}
Define the fluctuations of the Fermi surface as
\eqn\zhg{ \epsilon_\pm(t \mp q) = \mu + \delta \epsilon_\pm(t \mp q).}
$\delta \epsilon_\pm$ can be expanded in modes using \zha\ and \zhe.
Setting $m=0$ and  $n=ik$ ($k$ real and positive) in \zl, 
and substituting in the mode
expansion gives
\eqn\zhh{\eqalign{ a_k^\dagger = (\half \mu)^{-ik}
\sum_{n=1}^\infty {1  \over n!}\left({i \over
\sqrt{2\pi}\mu}\right)^{n-1}& {\Gamma(1-ik) \over \Gamma(2-n-ik)}\cr
&
\!\!\!\!\!\!\!\!\!\!\!\!\!\!\!\!\!\!\!\!\!\!\!\!\!\!\!\!\!
 \int_{-\infty}^0 \! dk_1 \ldots dk_n
(a^\dagger_{k_1} - a_{k_1}) \ldots  (a^\dagger_{k_n} - a_{k_n})\delta
(\pm |k_1| \pm \cdots \pm |k_n| -k).}}
The notation is somewhat schematic:  expanding out the string
of operators, each argument $\pm |k_i|$ of the delta function
comes with a plus sign if
the term contains $a^\dagger_{k_i}$, and a minus sign if it contains
$a_{k_i}$.

The result \zhh\ allows us to express any collection of incoming rightmoving
fluctuations in terms of outgoing leftmoving fluctations, and so yields
the S-matrix.
 The derivation of \zhh\ treated the creation/annihilation operators
as classical objects, neglecting operator ordering issues.  It is not
hard to check that the tree level S-matrix is independent of the choice
of ordering in \zhh, and we will find it convenient to use normal
ordering.

It is important to note that  the modes $a_k$ of the collective field
$S$  are nontrivially related to those of 
the tachyon as defined by the worldsheet CFT,
\eqn\fn{ [a_k]_{ws} = -i (4\mu)^{-i{k\over 2}}
{\Gamma(ik) \over \Gamma(-ik)} a_k.}
  Since the operators are related by pure phases for real $k$, these
terms only affect probabilities 
for  processes involving superpositions of different $k$.

\newsec{Rolling tachyon states}

Besides smooth fluctuations of the Fermi surface, another class of
states consists of exciting a single fermion. 
It is now understood 
\refs{\McGreevyKB,\MartinecKA,\McGreevyEP,\TakayanagiSM,\DouglasUP,\GaiottoYF}
that a single fermion on the top of the inverted harmonic oscillator
potential corresponds to 
a $D0$-brane localized in the strong coupling region.  A state in
which the fermion rolls down the potential 
corresponds to a rolling of the open string tachyon on the $D0$-brane.
At late times, as the fermion moves into 
the spatial region occupied by the Fermi sea, the state is best
described in closed string language as an outgoing 
pulse of radiation.   As shown in \KlebanovKM, there is a very precise
relation between the single fermion state and 
the profile of the outgoing pulse of closed string radiation.

Bosonization provides a dictionary between single fermion states and
collective field states \SenguptaBT\GrossST. We now review this 
dictionary in the region of large negative $x$; the full formulas are
found in \KlebanovQA.   Starting from the $\psi$ 
field appearing in \zb, we change variables as
\eqn\ya{\psi(x,t) = {1 \over x} e^{-i\mu t +{i\over 2}x^2} \psi_L(x,t)
+ {1 \over x} e^{-i\mu t -{i\over 2}x^2} 
\psi_R(x,t) }
which amounts to stripping off the WKB part of the wavefunction for
 large negative $x$.   Define $q$ as in \zhb\ 
 and substitute into the Hamiltonian,  keeping  only terms which
 survive for large negative $q$, to obtain 
\eqn\yb{ H= \int_{-\infty}^\infty \! dq \, \left[i \psi^\dagger_R \p_q
 \psi_R-i \psi^\dagger_L \p_q \psi_L\right].} 
This is the Hamiltonian of a relativistic fermion. Now bosonize as
\eqn\yc{\eqalign{ \psi_R & = {1 \over \sqrt{2\pi}} :\exp
\left[i\sqrt{\pi}\int^q (\pi_S-\p_q S)dq'\right]:\cr 
 \psi_L & = {1 \over \sqrt{2\pi}}
:\exp \left[i\sqrt{\pi}\int^q (\pi_S+\p_q S)dq'\right]:}}
 The bosonic field $S$ is the
same as the  field which appeared earlier in \zha.

Single fermion states are obtained by acting with $\psi_{L,R}$ on the
vacuum $|0\rangle$.  Of course, in our 
context the relevant ground state consists of the filled Fermi sea.
However, if we consider wavepackets which 
have very small overlap with states of the Fermi sea, we can
effectively consider acting on the zero particle 
ground state $|0\rangle$. Such states are
\eqn\yd{\eqalign{ \psi_R(q,t)|0\rangle  & = {1 \over \sqrt{2\pi}} \exp
\left[-2i\sqrt{\pi}\int_{0}^\infty \! {dk 
\over \sqrt{8\pi^2 k^2}}~ a_k^\dagger e^{-i(kq-|k|t)}\right]|0\rangle \cr
 \psi_L(q,t) |0\rangle & = {1 \over \sqrt{2\pi}} 
\exp \left[2i\sqrt{\pi}\int_{-\infty}^0 \! {dk \over \sqrt{8\pi^2
k^2}} ~a_k^\dagger 
e^{-i(kq-|k|t)}\right]|0\rangle}}

Classically, the fermions move along trajectories obeying the
equations of motion.  Trajectories with a turning 
point at $t=0$ are
\eqn\ye{ x(t) = - \lambdah \cosh t, \quad \lambdah=\sin \pi \lambda.}
%
%A trajectory at the Fermi surface corresponds to $\lambdah = \sqrt{2\mu}$.
  At early and late times, the
trajectories become relativistic
\eqn\yf{t\rightarrow \pm \infty:\quad   q(t) = \mp t -\ln {\lambdah\over 2}.}
Therefore, at early and late times we can write
\eqn\yg{ \psi_{R} = \psi_R(t-q), \quad \psi_{L} = \psi_L(t+q).}
Asymptotic states can be obtained by acting with either $\psi_R$ or
$\psi_L$ in the region of large negative $q$. 
We can choose to work in terms of the $\psi_R$ states, since the
$\psi_L$ states are related to these by 
reflecting off the potential.  It turns out to be convenient to write
the incoming state as 
\eqn\yh{ \psi_R(t-q+\ln {\mu\over 2})|0\rangle, \quad t-q=-\ln
{\lambdah \over 2}.} 
The $\ln {\mu\over 2}$ shift is convenient since it will cancel the
prefactor in \zhh. 

According to the bosonization \yd, the state \yh\ is
\eqn\yi{\eqalign{\sqrt{2\pi}\psi_R(t-q+\ln {\mu\over 2})|0\rangle &=
\exp \left[-2i\sqrt{\pi}\int_{0}^\infty \! 
{dk \over \sqrt{8\pi^2 k^2}}~ a_k^\dagger e^{ik(t-q+\ln {\mu\over
2})}\right]|0\rangle \cr & =\exp 
\left[-2i\sqrt{\pi}\int_{0}^\infty \! {dk \over \sqrt{8\pi^2 k^2}}~
a_k^\dagger \mu^{ik} e^{-ik\ln 
\lambdah}\right]|0\rangle}}

The state \yi\ can be viewed either as a single incoming fermion  (the
``open string interpretation"), or as an 
incoming pulse of tachyons (the ``closed string interpretation".)  We
can now use our tree level S-matrix result 
\zhh\ to reexpress it in terms of  outgoing states as
\eqn\yk{ \eqalign{ &  \exp\left[ 2i\sqrt{\pi} \sum_{n=1}^\infty {1
\over n!} \int_0^\infty \!  {dk \over 
\sqrt{8\pi^2 k^2}} \int_{-\infty}^0 \! dk_1 \cdots dk_n
e^{i\theta(k)} \left({i \over \sqrt{2\pi} 
\mu}\right)^{n-1} {\Gamma(1-ik)\over \Gamma(2-n-ik)} \right. \cr &
\quad\quad\quad\quad 
\quad\quad\quad\quad\quad\quad\quad
 :(a^\dagger_{k_1}-a_{k_1})\dots (a^\dagger_{k_n}-a_{k_n}):
\delta(\pm |k_1| \pm \cdots \pm |k_n| - k )\Bigg] |0\rangle}}
with
\eqn\yl{ e^{i\theta(k)} =-2^{ik} e^{-ik\ln \lambdah}.}
\yk\ represents a particular state of outgoing tachyons, and whose
expansion can be matched against contributions 
in  string perturbation theory.

As an expansion in $g\sim 1/\mu$, if we keep just the lowest $n=1$ term we find the state
\eqn\ym{  \exp\left[ 2i\sqrt{\pi}\int_{-\infty}^0 \!
  {dk_1 \over \sqrt{8\pi^2 k_1^2}}
e^{i\theta(|k|)} a^\dagger_{k_1} \right] |0\rangle.}
As was shown in \KlebanovKM, 
this state agrees with that produced by the disk one-point function,
or more precisely, by 
the sum over any number of disks with one vertex operator inserted on
each. To get agreement, the time part of the 
CFT should include a boundary interaction $\lambda \cosh t$, and the
zero mode should be integrated using the 
Hartle-Hawking contour extending to $t =
+i\infty$.\foot{Alternatively, one can use the boundary interaction 
$\lambda e^t$.  We will think in terms of the $\lambda \cosh t$
interaction since it seems to correspond more 
naturally to our setup at early times.}  One feature of the
Hartle-Hawking contour is that it restricts us to 
computing just the production of closed string states and not their
absorption, since convergence of the zero mode 
integral requires vertex operators to behave as $e^{i\omega t}$ with
$\omega>0$. 

If we only keep terms with all creation operators we get
\eqn\yn{  \exp\left[ 2i\sqrt{\pi}  \sum_{n=1}^\infty {1 \over n!}
\int_{-\infty}^0 \! dk_1 \cdots dk_n 
e^{i\theta(|k|)} \left({i \over \sqrt{2\pi} \mu}\right)^{n-1}
 {\Gamma(1-ik)\over \Gamma(2-n-ik)} 
 a^\dagger_{k_1} \cdots a^\dagger_{k_n}
\right] |0\rangle. }
This has the correct form to arise from the sum of disk diagrams with
any number of tachyon vertex operators. The 
disk amplitudes should be evaluated using the same Hartle-Hawking
contour as above. 

As will be shown in more detail later, the remaining terms in the
expansion of \yk\ come from worldsheets with 
multiple boundaries
--- the annulus and so on. We should emphasize that the only
approximation we have made was to treat the closed 
strings (the collective field) classically, which means that we are
correctly including quantum effects due to 
open strings. Of course, this is very familiar from the AdS/CFT
correspondence, where we are used to saying that 
classical nonlinear closed string effects are dual to quantum open
string effects.

\newsec{Field coupled to classical source}

To make a precise connection between string amplitudes and the state
\yk, we need to recall a few basic facts 
regarding the states produced by classical sources.  Start from
\eqn\xa{ S = \half \int \! d^2x \left[ \dot{\phi}^2 - (\phi')^2 +
J\phi\right].} 
The field equations are solved in terms of the retarded propagator as
\eqn\xb{ \phi_{out} = \phi_{in} + \int \! d^2 x' G_{ret}(x-x')J(x')}
\eqn\xc{G_{ret}(x) = \int \! {d^2 k \over (2\pi)^2} {e^{i(kx-\omega
t)} \over -(\omega+i\epsilon)^2 +k^2} } 
We define the mode expansions  as
\eqn\xd{\eqalign{ \phi_{in} & = \int_{-\infty}^\infty \! {dk \over
\sqrt{8\pi^2 k^2}} \left[ b_k e^{i(kx-|k| t)} + 
b_k^\dagger e^{-i(kx-|k| t)}\right] \cr \phi_{out} & =
\int_{-\infty}^\infty \! {dk \over \sqrt{8\pi^2 k^2}} 
\left[ a_k e^{i(kx-|k| t)} + a_k^\dagger e^{-i(kx-|k| t)}\right]}}
It then follows that the in vacuum is a coherent state when expressed
in terms of out operators, 
\eqn\xe{ |0_{in}\rangle =\exp\left(i \int {dk \over \sqrt{8\pi^2 k^2}}
\tilde{J}(|k|,k)a_k^\dagger\right)|0_{out}\rangle }
with
\eqn\xf{\tilde{J}(\omega,k)= \int \! d^2x \,J(x)e^{-i(kx-\omega t)}.}

Comparing \xe\ with \ym, we can read off the source corresponding to
the disk one-point function; we find 
\eqn\xg{ \tilde{J}(|k|,k) =2\sqrt{\pi}e^{i\theta(|k|)},\quad (k<0).}

\subsec{Generalization}

Now consider
\eqn\xh{S= \half \int\!d^2 x \left[ \dot{\phi}^2- (\phi')^2\right]+
\int \! d^2x_1 \cdots d^2x_n \, 
J(x_1,\ldots,x_n) \phi(x_1)\cdots \phi(x_n).}
In first order perturbation theory we have
\eqn\xi{\eqalign{ |0_{in}\rangle & = e^{-i\int H(t) dt}
|0_{out}\rangle = |0\rangle + i\int \! d^2x_1 \cdots 
d^2x_n \, J(x_1,\ldots,x_n) \phi(x_1)\cdots \phi(x_n)|0_{out}\rangle
\cr & = |0_{out}\rangle+i \int \! {dk_1 \over 
\sqrt{8\pi^2 k_1^2}} \cdots {dk_n \over \sqrt{8\pi^2 k_n^2}} \,
\tilde{J}(|k_1|,k_1; \cdots ; |k_n|,k_n)\, 
a_{k_1}^\dagger \cdots a_{k_n}^\dagger|0_{out}\rangle }}
where
\eqn\xj{ \tilde{J}(k_1,\cdots ,k_n) =\int\! d^2 x_1 \cdots d^2 x_n \,
J(x_1,\cdots, x_n)e^{-i(k_1\cdot x_n+\cdots 
+ k_n \cdot x_n)}.}
Comparing this to \yn\  we find
\eqn\xk{\tilde{J}(|k_1|,k_1; \cdots ; |k_n|,k_n) = 2\sqrt{\pi}
e^{i\theta(k)} {\prod_{i=1}^n \sqrt{8\pi^2 k_i^2} 
\over \sqrt{8\pi^2 k^2}}{1 \over n!}\left({1 \over
\sqrt{2\pi}\mu}\right)^{n-1} {\Gamma(1-ik) \over 
\Gamma(2-n-ik)}.}
Here $k_i <0$ and $k = \sum_i |k_i|$.  So \xk\ is our prediction for
the disk n-point functions.

\newsec{Comparison with string perturbation theory}

As was shown in \KlebanovKM, the result \xg\ agrees with the disk
one-point function, after taking into account the additional 
leg-pole factors \fn.   In particular, we consider the one-point
functions of the normalizable vertex operators 
\eqn\wa{ V_k = e^{(2+ik)\phi-i|k|t}.}
The Liouville part is described by the Zamolodchikov boundary state 
\ZamolodchikovAH,
\eqn\wb{ \langle e^{(2+ik)\phi} \rangle = {2 \over \sqrt{\pi}}i
\sinh(\pi k) \mu^{-i{k \over 2}} {\Gamma(ik) \over 
\Gamma(-ik)} }
and the time part is described by the boundary interaction $\lambda
\cosh t$ with Hartle-Hawking contour, 
\eqn\wc{ \langle e^{i|k|t}\rangle = {\pi e^{- i|k| \log {\lambdah}}
\over \sinh(\pi |k|)}.} 
 Combining these gives
\eqn\wd{ \langle V_k \rangle = 2 \sqrt{\pi}   e^{-i|k| \log
{\lambdah}} e^{i\delta(k)}} 
with
\eqn\we{ e^{i\delta(k)} = i\,{\rm sgn}(k) \mu^{-i{k \over 2}}
{\Gamma(ik) \over \Gamma(-ik)}.} 
\wd\ agrees with \xg\ modulo a $\lambda$ independent phase, which is
attributed to the leg pole factor \fn.

\subsec{Generalization}

By the same logic, we can relate our general result \xk\ to the
general disk amplitude with any number of outgoing 
tachyon vertex operators.  Modulo the leg pole factors, we then  get a
prediction for the disk amplitudes of the 
rolling tachyon times Liouville boundary state:
\eqn\eh{ \langle V_{k_1,\cdots, k_n}\rangle = 2\sqrt{\pi}
{\prod_{i=1}^n \sqrt{8\pi^2 k_i^2} \over \sqrt{8\pi^2 
k^2}}{1 \over n!}\left({1 \over \sqrt{2\pi}\mu}\right)^{n-1}
{\Gamma(1-ik) \over \Gamma(2-n-ik)}.} 
The precise statement is that these are the amplitudes for outgoing
particles evaluated using the Hartle-Hawking 
contour.

\subsec{Amplitudes with multiple boundaries}

Now we return to the issue of the remaining states in the expansion of
 \yk.   For illustration, consider working 
 to first order in $g \sim 1/\mu$.  At this order two amplitudes
 contribute: the disk with two vertex operators, 
and the annulus with one vertex operator.\foot{Amplitudes with no
 vertex operators of course do not contribute to 
the closed string state.}  The claim is that the annulus amplitude
 arises upon normal ordering \yk. In particular, 
we can write
\eqn\ma{ e^{A +g B} = e^A \left\{  1+ g\left( B + {1 \over 2!} [B,A]
+ {1 \over 3!} \big[[B,A],A\big] + \cdots \right) \right\} + {\rm O}(g^2).}
% 
%\ma\ is  valid to order in $g$ provided that (as is true in our case)
%$\Big[[B,%A],A\Big]=0$. 
 In our case
\eqn\mb{\eqalign{ A& = 2i\sqrt{\pi} \int_{-\infty}^0 \!{dk \over
\sqrt{8\pi^2 k^2}} e^{i\theta(|k|)} a^\dagger_k 
\cr gB&= {i \over \sqrt{2} \mu} \int_{-\infty}^0 \! {dk_1 \, dk_2\over
\sqrt{8\pi^2}} \, \left( 
e^{i\theta(|k_1|+|k_2|)} a^\dagger_{k_1} a^\dagger_{k_2} +2
e^{i\theta(|k_1|-|k_2|)}\Theta(|k_1|-|k_2|) a^\dagger 
_{k_1} a_{k_2}\right) }}
We then find
\eqn\me{ g[B,A] = -i {(2\pi)^{3/2} \over \mu} \int_{-\infty}^0 \! dk_1
k_1^2 e^{i\theta(|k_1|)} a_{k_1}^\dagger, } 
and the remaining nested commutator terms in \ma\ vanish.
From this we read off the annulus one-point function to be
$\half(2\pi)^{3/2} k^2$ (modulo a phase factor). 

The annulus amplitude amplitude grows rapidly in $k$, and the
corresponding state obtained from \me\ is an 
infinite energy, highly non-normalizable, state.  This is not a surprise: 
this behavior is obtained already at the 
level of the disk, and the two boundaries of the annulus essentially squares this divergence.  As discussed in \KlebanovKM,
 these divergences have a simple origin:
our incoming state was taken to be a single fermion with a definite 
position, but such a state has infinite energy quantum mechanically.  
A more accurate treatment replaces the fermion state by a localized 
wavepacket;  the wavepacket then cuts off the large $k$ divergences.  
Note that this also  makes it clear that the divergence in the energy of the
outgoing state is not somehow cured by 
including all terms in the expansion of \yk, since it is obvious that our
single, perfectly localized, fermion state has an infinite energy at
early times, and energy is conserved.

To further illustrate that \me\ yields an annulus amplitude, consider
replacing the single rolling fermion by $N$ 
of them.   Since this corresponds to $N$ unstable D-branes, we should
find that the disk with two vertex operators 
is proportional to $N$, while the annulus with one vertex operator is
proportional to $N^2$.   At our level of 
approximation, the $N$ fermion state is obtained by simply inserting
an $N$ in the exponent of \yi.   This then 
 modifies \ma\ to
\eqn\maa{ e^{N(A +g B)} = e^{NA} \left\{  1+ g\left( NB + \half
N^2[B,A] \right) \right\} +  {\rm O}(g^2),} 
which immediately leads to the correct $N$ behavior.

Amplitudes with more than two boundaries are similarly obtained by 
considering higher order terms in the exponent of \yk; i.e. expanding 
an expression of the form $e^{N( A+ g^{n-1} B)}$ to first order in
$g^{n-1} B$.   
Using \ma, the series starts with a term representing $n$ vertex operators
on the disk.  Each addition of a commmutator with $A$ removes a vertex
operator (since one less creation operator appears) and adds a boundary
(since the amplitude is proportional to one higher power of $N$.)
If we hold $N$ fixed, then we cannot really justify keeping amplitudes
with arbitrary number of boundaries without also including worldsheets
with handles.  In particular, the torus amplitude appears
at the same order in $g$ as the annulus.  In principle, the 
contribution of handles could be incorporated by replacing \yk\ by the
full perturbative tachyon S-matrix.  Alternatively, as in AdS/CFT,
if we take $g\rightarrow 0$ and $N\rightarrow \infty$ with $gN$ fixed, 
then the effect of handles is suppressed compared to adding any number 
of boundaries.   

The complete expansion of \yk\  also includes the  contributions from
disconnected worldsheets, since it describes the complete closed string
state.

\newsec{Static unstable D-branes in the matrix model}

Classically, a fermion placed at the top of the inverted harmonic
oscillator potential can stay there forever. In the two dimensional
string theory this corresponds to the fact that one can construct a
boundary state corresponding to an eternal unstable D-brane. 
The boundary state is constructed by tensoring  the 
$(m,n)=(1,1)$ boundary state of Zamolodchikov and Zamolodchikov 
  with a
Neumann boundary state for the free time directions $X_0$. Quantum
mechanically, a localized wavefunction will spread and the unstable
brane will have a finite lifetime. In the string theory the
instability of the static unstable D0 brane manifests itself in the
appearance of an imaginary part in the annulus partition function
\MarcusVS\CrapsJP.  

It is an interesting
question to ask whether such an unstable brane will modify
the classical closed string scattering.  From the worldsheet perspective
one again expects to find corrections from disk amplitudes. 
 But on the matrix model side our preceding analysis 
does not directly apply since it heavily used the bosonization of the
fermion in the asymptotic region, whereas here we want to keep the fermion 
at the origin. 
In this section we attempt to address this question using a modified version
of collective field theory.  Surprisingly, we will in fact   find 
vanishing corrections corresponding to disk amplitudes. 

\subsec{Modified collective field theory}

In the singlet sector the matrix model action \za\  reduces to the
following action for the eigenvalues

\eqn\lgrad{L= \sum_i \Big( {1\over 2} (\partial_t \lambda_i)^2 - {1\over 2}
\sum_{j\neq i} {1\over (\lambda_i -\lambda_j)^2} - V(\lambda_i) \Big).}
In \lgrad\ the eigenvalues are treated as bosonic with a repulsive potential.
Das and Jevicki \DasKA\  introduced a collective field to describe the dynamics
of the eigenvalues in the large N limit
\eqn\colla{\partial_x \phi(x,t)= \sum_{i} \delta( x- \lambda_i(t)).}
The dynamics of the collective field $\phi$ is described by the
following Lagrangian
\eqn\collb{L=\int dx \Big( {1\over 2} {\partial_t \phi \partial_t \phi
\over \partial_x\phi}- {\pi^2\over 6} (\partial_x \phi)^3 - ( V(x) -
\mu_F)\partial_x \phi\Big).} 
In the double scaling limit the potential is given by $V(x)={1\over
2}(V_0- 
x^2) $ and one takes $N\to \infty$, 
$\bar \mu = V_0-\mu_F\to 0$,  keeping $N\bar \mu = \mu $ fixed. The Lagrangian
\collb\ becomes then 
 \eqn\collc{L=\int dx \Big( {1\over 2} {\partial_t \phi \partial_t \phi
\over \partial_x \phi}- {\pi^2\over 6} (\partial_x \phi)^3 + ({1\over
2}x^2 
-\mu) \partial_x \phi\Big).}
Note that this Lagrangian  can be derived from the Hamiltonian \zg\ by
defining $p_{\pm}= -P_\phi\pm \pi \partial_x \phi$ and eliminating
the momentum $P_\phi$ via a Legendre transformation. The string
coupling constant is related to  the height of the double scaled
potential by $\mu=1/g$.

A variation on the collective field theory of Das and Jevicki was
developed by Brustein et al. \BrusteinBI\BrusteinMY\foot{An
alternative description of single eigenvalue tunneling was developed
in \DharCS, based on the formalism of \DharHR.}.   One splits a 
single eigenvalue $\lambda_0(t)$ from the    
collective field $\phi(x,t)$ and treats it separately. This is justified for
eigenvalue distributions  where the single eigenvalue is away from the dense
region of the Fermi sea, i.e. $|\lambda|\ll {1\over g}$. The dynamics
of the filled Fermi sea is again described by the collective field $\phi(x,t)$.
The action for the coupled system is 
\eqn\collh{L= {1\over 2} (\partial_t \lambda_0)^2 + {1\over 2}
\lambda_0^2 - \int dx {\partial_x \phi\over (x-\lambda_0)^2}+\int dx
\Big( {1\over 2} {\partial_t \phi \partial_t \phi 
\over \partial_x \phi}- {\pi^2\over 6} (\partial_x \phi)^3 + ({1\over
2}x^2 
-{1\over g}) \partial_x \phi\Big).}
In order to disentangle the dynamics of the single eigenvalue and the
collective field   it is useful to perform the following rescaling
$\phi= g^{-1} \hat \phi$, $x= g^{-{1\over 2}} \hat x$ and $
\lambda = g^{-{1\over 2}}\hat \lambda$. 
The action \collh\ becomes
\eqn\collhb{L= {1\over g} \Big(  {1\over 2}(\partial_t \hat
\lambda_0)^2 + {1\over 2} 
\hat \lambda_0^2\Big) - \int d\hat x {\partial_{\hat x} \hat \phi\over
(\hat x-\hat \lambda_0)^2}+{1\over g^2}\int d\hat x
\Big( {1\over 2} {\partial_t \hat \phi \partial_t \hat \phi 
\over \partial_{\hat x} \hat \phi}- {\pi^2\over 6} (\partial_{\hat x}
\hat \phi)^3 + ({1\over
2}\hat x^2 
-1) \partial_{\hat x} \hat \phi\Big).}
%The form of the action is very suggestive and it is tempting to
%view the action 
The form of 
\collhb\ suggests an interpretation as an open-closed string field
theory action.  
The part which is of order
$1/g$ can be interpreted as the action for the open string degree of
freedom associated with the unstable D-brane. The part which is of
order $1/g^2$ is the Das-Jevicki collective field action and
corresponds to the action for the closed strings.  
 The coupling between
open and closed strings comes at order $g^0$ and 
in the limit $g\to 0$ becomes unimportant. Furthermore the $g\to 0$
 limit corresponds to the limit where $\lambda_0$ and $\phi$
can be treated as classical fields.

A simple solution of the decoupled equation is given by 
\eqn\solsim{\lambda_0(t)= a_1 \cosh(t)+a_2 \sinh(t), \quad\quad
\partial_x\phi={1\over \pi}\sqrt{x^2-{2\over g}},}
corresponding to a rolling eigenvalue and a static Fermi sea. It is
interesting to analyze what happens if $g$ is small but nonzero. When
the eigenvalue $\lambda$ is of order $1/g^{1\over 2}$ the interaction term
becomes important. For the rolling tachyon \solsim\ this happens at a
time $t= -\log g^{1\over2}$. One might be tempted to argue that this
implies that there is a strong interaction between the eigenvalue and
the Fermi sea, which starts at $x^2=2/g$. However this is not clear,
since the derivation of the action \collh\ assumed that the single
eigenvalue is well separated from the Fermi sea, so the action
\collh\ might not be a good description of the actual dynamics in this
case.

\subsec{Scattering from a static D-brane}
Instead of the rolling tachyon,  in the classical limit one can
consider a solution where the eigenvalue sits on top of the inverse
harmonic oscillator potential $\lambda(t)=0$ for all $t$.
 The equation of motion for the collective field then becomes
\eqn\collj{
\partial_t\left({\partial_t \phi \over \partial_x \phi}\right)- \partial_x
\Big\{ {1\over 2} \left({\partial_t \phi\over \partial_x \phi}\right)^2
+ {\pi^2\over 2} (\partial_x\phi)^2 - {1\over 2}x^2+{1\over g}+
{1\over x^2}\Big\} 
=0. }
The interaction term modifies the static solution 
\eqn\statsola{\partial_x \phi= {1\over \pi}\sqrt{ x^2 -{2\over g}-
{2\over x^2}},} 
which is valid for $x^2> 2/g$.  Note that at large $x$ the corrections
to the standard static solution 
\zf\ are of order $1/x^3$ and subleading. A possible interpretation
is that a localized D-brane near $x=0$ only has a weak backreaction on
the fields in the weak coupling region at $x=-\infty$.

The  Das-Jevicki collective field is related to the  description of
the dynamics of the Fermi sea of Polchinski
by $\partial_x \phi= {1\over 2\pi}(p_+-p_-)$. This implies that the equation
of motion for $p_\pm$ get modified to
\eqn\dbrana{\partial_t p_\pm = x+ {2\over x^3} - p_\pm \partial_x
p_\pm .}
In the classical limit the motion of the fermions is described by an
incompressible fluid moving in a potential. From \dbrana\ it is clear
that the eigenvalue at $\lambda_0=0$ produces a small modification of
the Hamiltonian which governs the motion of points in the phase space,
\eqn\dbranb{H={1\over 2} p^2 -{1\over 2} x^2 +{1\over  x^2} +\mu.}
As closed string excitations are represented by small ripples in the
Fermi sea whose turning point is at $x^2\sim 1/g$, the extra term in
\dbranb\ is only a small perturbation.
The equations of motion following from \dbranb\ are hence
\eqn\hamc{{d^2 \over dt^2}x(t) -x(t) - {2\over x(t)^3}=0.}
It is interesting that the equation of motion
\hamc\ is one of the few modifications of the (inverted) harmonic
oscillator which can be solved exactly:
\eqn\hamb{x(t)= a\sqrt{\cosh^2(t-\sigma)+ b}, \quad 
b =-{1\over 2}( 1- \sqrt{1- {8\over a^4}}).}
%where the constant $b$ is given by the root of a quadratic equation
%\eqn\hamx{ }
We have chosen the root for which \hamb\ goes over to the solution
for the inverted harmonic oscillator in the limit $g\to 0$. 
We now follow the arguments of Polchinski to derive the classical
scattering from the time delay.
The time delay for the motion from a given $x$ and back is calculated
using \hamb,
\eqn\hamf{t'-q =t+q +  \ln({a^2\over 4}),} 
up to terms which vanish exponentially at late and early times
respectively. Using the relation $\epsilon_{\pm}= \pm (p\pm x)x$ one
finds
\eqn\hamh{\epsilon_{-}(t+q)= {a^2\over 2}\sqrt{1-{8 \over a^4}}.}

The relation  $\epsilon_-(t+q)=\epsilon_+(t'-q)$ can be  reexpressed
using \hamf\ 
and \hamh\ to produce a nonlinear  relation between 
 incoming and outgoing waves.
\eqn\smata{\epsilon_-(t+q) = \epsilon_-\big( t+q  + \ln( {1\over 2}
\sqrt{\epsilon_-(t+q)^2 +2})\big)  .} 
Expanding $\epsilon_+(t-q)= {1\over  g}+ \delta_+(t-q)$,
$\epsilon_-(t+q)={1\over  g}+
\delta_-(t+q)$  as in \zhg\ and using the formulas \zha\ and \zhe\ one can
calculate the S-matrix. It is easy to see that the formula for the
time delay \smata\ 
and hence for the  S-matrix are modified at order $g^2$. 

This is a puzzling feature since
from string perturbation theory one would  have
expected that the corrections are of order $g$,  coming from the disk 
versus  sphere diagrams.  This result can be traced back to the fact
that the interaction term in \collhb\ comes at order $g^0$ instead of
order $1/g$. It would be interesting to understand this fact better
from the second quantized fermion point of view.  The basic puzzle is that
a fermion on the top of the potential has only an exponentially small
overlap with states of the Fermi sea, and so would not seem to 
affect the perturbative scattering amplitudes.

A possible interpretation is that the 
backreaction on the closed string background caused by the presence of
the fermion on the top of the potential is weaker than
one might have expected. An indication that this is the case comes
from  considering the boundary state representing a static D-brane
in 2 dimensions.  The vertex
operator for on shell 'massless' tachyon $V_k$ is  given by \wa.
The one point function on the disk of the Liouville primary $ U_k =e^{(2 +
ik)\phi}$ is given by \wb.
Note that the Neumann boundary conditions on the $X_0$ enforce $k=0$ 
by momentum conservation. Hence in contrast to the rolling tachyon
boundary state discussed in section five, the only physical state appearing in
the boundary state is $V_k$ with $k=0$.
 It follows that the
one point function of this state is zero because of the $\sinh(k)$
factor in \wb.   

Our results further seem to imply the vanishing of all
disk amplitudes with on-shell closed string vertex operators.  This 
conclusion is in fact consistent with the T-dual version of our
setup, studied in  \AlexandrovNN
\foot{We thank David Kutasov for explaining this to us.}  It would be nice
to gain a better understanding of these vanishing amplitudes from 
the worldsheet point of view.

\medskip

\newsec{Discussion}

We have discussed closed string amplitudes in the presence of unstable
D0-branes. 
In the case of the rolling tachyon background, we were able to relate the 
known sphere amplitudes for closed string tachyons to the sum of
planar amplitudes 
with any number of boundaries.  This represents a  relation of the
type one is familiar with from AdS/CFT duality, but here we are able
to explicitly  
compute
quantities on both sides of  the duality. Of course, it would be
interesting to  
reproduce these results using continuum methods.
  Since the matrix model results 
correspond to worldsheet amplitudes using the Hartle-Hawking time contour,
we were restricted to considering only outgoing vertex operators.  It would
be interesting to generalize this, as well as to be able to treat the
real time 
contour.    

For black holes in $AdS_5$ there is a phase transition at high
temperatures which liberates the underlying open string degrees of
freedom \WittenZW.  It is interesting that we found  a hint of the non
Abelian
structure of the open strings in the combinatorics of boundaries in
the matrix model scattering. This suggests the  possibility
to create a two dimensional black
hole by bringing many unstable D0-branes together. It would be
interesting to relate  this idea  to other proposals for a matrix model
for the two dimensional black hole \KazakovPM.

Scattering amplitudes for the static D-brane are more surprising.  We employed
a modified version of the collective field formalism in which one separates
out a single eigenvalue from the continuum. This results in a deformed 
Fermi sea, and corresponding corrections to tachyon amplitudes.   But since
these corrections arise at order $g^2$,  whereas disks contribute
at order $g$,  we find that the disk amplitudes vanish, in agreement 
with previous results \AlexandrovNN.   
\bigskip
\bigskip
\bigskip
{\bf Acknowledgements}
\medskip
\noindent The work of MG is supported in part by NSF grant 0245096,
and the work 
of PK  is supported in part by NSF grant 0099590. Any opinions,
findings and conclusions expressed in this material are those of the
authors and do not necessarily reflect the views of the National
Science Foundation. 
\listrefs
\end